# The Zero-n Gap Soliton

Ravi S. Hegde, Herbert G. Winful

*Department of Electrical Engineering and Computer Science, University of Michigan, Ann Arbor, MI 48109*

**Abstract**. Periodic structures consisting of alternating layers of positive-index and negative-index materials possess a novel band gap at the frequency at which the average refractive index is zero. We show that in the presence of a Kerr nonlinearity, this zero-n gap can switch from low transmission to a perfectly transmitting state, forming a nonlinear resonance or gap soliton in the process. This zero-n gap soliton is omnidirectional in contrast to the usual Bragg gap soliton of positive –index periodic structures.

Periodic dielectric structures possess stopbands in frequency within which an electromagnetic wave undergoes coherent Bragg scattering and is totally reflected by the structure. In 1979, Winful et al showed that the inclusion of an intensity-dependent refractive index makes it possible to tune the stop band and thus switch the structure from a highly reflecting state to a totally transmitting state [1]. In the totally transmitting state, the structure exhibits a nonlinear resonance with a field distribution that peaks at the center [2]. This nonlinear resonance structure associated with the Bragg gap is known as a gap soliton [3] and has been the subject of intense studies [4]. Recently, a completely new kind of band gap has been discovered in structures where the average refractive index vanishes at certain frequencies [5-9]. One way to achieve such a zero-n gap is to create a periodic structure with alternating layers of positive index materials and negative index materials (also termed "left handed materials" by Veselago [10]). The zero-n gap differs from the usual Bragg gap in that it is invariant to scale length and relatively insensitive to disorder and input angle [5, 6]. The effect of nonlinearity on this zero-n gap has only recently been discussed and phenomena such as hysteresis and bistability have been predicted [11, 12].

In this Letter we show that a novel kind of gap soliton associated with the zero-n gap exists in periodic structures with alternating positive and negative index layers. We call this the "zero-n gap soliton" and show that it is relatively insensitive to input angle unlike the usual Bragg gap soliton. This raises the intriguing possibility of an omni-directional gap soliton that is also robust in the presence of disorder.

We consider a structure in which the positive index medium exhibits Kerr-type nonlinearity. The negative index medium is taken to be linear although nonlinearity in its electric and magnetic susceptibilities can be easily included. We ignore absorption in the structure, which can also be readily included within this formalism The periodic structure (as in Fig. 1) consists of N unit cells occupying the region z = 0, D, translationally invariant in the x-y plane and bounded on both sides by free space. Each cell is formed of two films film 1 and film 2 with thicknesses $d_1$ and $d_2$ mm, each less than the wavelength of the electromagnetic wave. Film 1 is a positive index material, while film 2 is a negative index or left handed material. The dielectric permittivity of film 1, described by Eq. (1) is taken to exhibit field intensity dependence. We ignore such effects for the magnetic permeability as done in [13].

$$\hat{\epsilon}_i(\omega, z) = \epsilon'_i(\omega, z) + \epsilon_{iNL}(|\mathbf{E}|^2) \quad (i = 1, 2) \quad (1)$$

The material constants for the left handed layer are dispersive and given (as a function of frequency f in GHz) by Eq. (2).

$$\epsilon'_2(f) = 1.6 + \frac{40}{0.81 - f^2},$$
$$\mu_2(f) = 1 + \frac{25}{0.814 - f^2} \quad (2)$$

For this particular choice of parameters, this material has a negative refractive index for $f < 5$ GHz. The implicit assumption in this formulation is that the negative index layer permits the use of an effective index for the frequency range of interest.

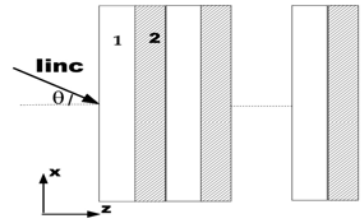

**FIGURE 1**. Schematic of the nonlinear periodic structure.

Adopting an approach similar to [14], we solve for TE polarized fields of the form,

$$E(x,z) = E(z)e^{ik_o(\beta x - ct)}$$
$$k_o = \omega/c$$
$$\beta = \sin(\theta) \quad (3)$$

where θ is the angle which the incident wave vector makes with the x axis. Now setting $\zeta = k_o z$. The electric field is in general governed by Eq. (4-5).

$$\frac{d^2 E}{d\zeta^2} + p^2 E + \mu_i \epsilon_{iNL}(|\mathbf{E}|^2) = 0 \quad (4)$$
$$p^2 = \mu_i(\epsilon_i'(\omega, z) - \beta^2) \quad (5)$$

The solutions for the most general case have to be obtained by numerical integration of eqn (3). In case of a Kerr-type nonlinearity, the $\epsilon_{iNL}(|E|^2)$ term in eqn. (1) can be replaced by $\gamma|E|^2$. To obtain the transmission of the structure we specify the E field and its derivative at the output face, successively integrate eqn (4) all the way to the input end. At the interfaces of two films we apply the following boundary conditions.

$$E_i = E_j$$
$$\mu_j \frac{dE}{d\zeta}\bigg|_i = \mu_i \frac{dE}{d\zeta}\bigg|_j \quad (6)$$

For N = 32, $d_1 = d_2 = 10$ mm, the zero-n gap is centered around 3.55 GHz where the following condition is satisfied.

$$\sqrt{\epsilon_1(f)}\sqrt{\mu_1(f)}d_1 = \sqrt{\epsilon_2(f)}\sqrt{\mu_2(f)}d_2$$

Inclusion of nonlinearity makes it possible to tune the gaps, resulting in bistability and the creation of gap solitons. Fig. 2 shows the transmission of the structure as a function of incident intensity for a frequency within the zero-n gap. The transmission is multivalued leading to hysteresis and bistability. Also notice that the hysteresis curves for different incident angles almost overlap at the high transmittance point. At the point marked A the transmission is nearly unity, signifying a transmission resonance. The spatial distribution of the magnitude of the E field associated with this transmission resonance is shown in Fig 3. This is the zero-n gap soliton. In the figure, the soliton is overlaid upon the periodic structure, with shaded regions representing the negative index material.

The zero-n gap exhibits several interesting features. It can be seen that for the zero-n gap soliton the maxima and minima of the intensity distribution occur at the interfaces between the layers. In the middle of the soliton where the intensity is high, the peaks begin to bifurcate and thus introduce a new set of minima. The overlaid graphic in Fig. 3 shows this feature. It can be seen that the minima of this soliton distribution do not go all the way to zero. A look at the phase variation of the E field inside the structure (Fig. 4) helps to understand this behavior. The phase does not build up to $\pi$ across a layer. Because there is not a $\pi$ phase shift, complete destructive interference does not occur between layers. It is also seen that at the interface between a positive index and negative index layer there is a change in the sign of the slope of the electric field. This is a result of the sign change in magnetic permeability as one crosses from a positive index to a negative index medium.

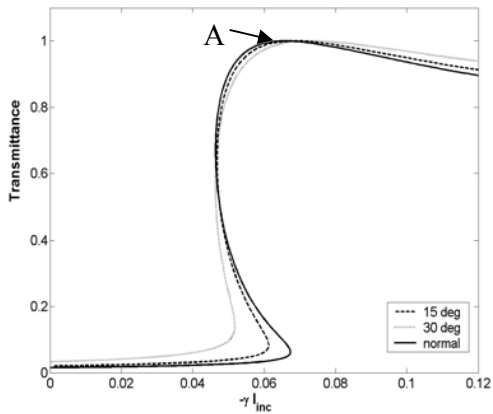

**FIGURE 2.** Hysteresis behavior of transmittance as a function of a defocusing $\gamma I_{inc}$ for detuning to the left of the zero-n gap (f = 3.51 GHz, N = 32.) for incident angles θ = 0°, 15° and 30°

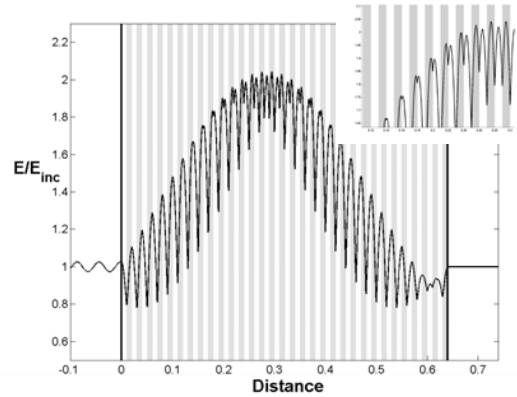

**FIGURE 3**. The Zero-n gap soliton, Spatial distribution of the E field magnitude (normalized w.r.t. $E_{inc}$) when a defocusing $\gamma I_{inc}$ = -0.064 results in a near unity transmittance (f = 3.51 GHz, N = 32.) at normal incidence. The overlaid graphic zooms in, to show the bifurcation observed at the top of the soliton.

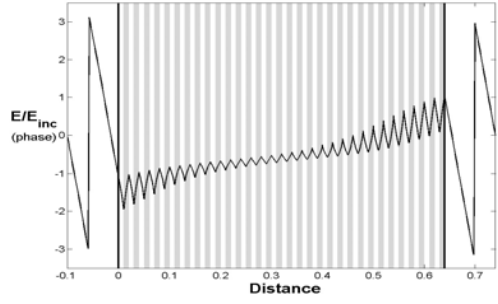

**FIGURE 4**. Spatial distribution of the phase of the E field for conditions as in Figure 3.

For comparison the gap soliton associated with the Bragg gap soliton is shown in Fig. 5. The frequency is chosen to lie in the Bragg gap. It is seen that the minima of the field pattern are nearly zero for the Bragg gap soliton. This is because the evolution of the phase across a bilayer is continuous and goes all the way to $\pi$.

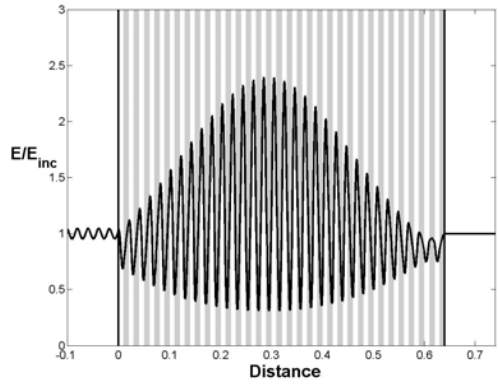

**FIGURE 5**. The Bragg gap soliton observed when a defocusing $\gamma I_{inc} = -0.027$ is used. (f = 7.15 GHz, N = 32.) at normal incidence.

For the same input intensity, the field distribution on the lower branch of the hysteresis curve for the zero-n gap is shown in Fig. 6. The minima in this case also does not go all the way to zero. This feature is again in contrast to the distribution for the corresponding Bragg gap case. This behavior can also be understood in terms of phase evolution.

It is of interest to see if the zero-n gap soliton exhibits some of the interesting features like ominidirectionality and robustness of the linear zero-n gap itself. The dependence of the zero-n gap soliton on incidence angle is shown in Fig. 7. For the same input intensity it persists for angular detuning as large as 30 degrees from the normal. Thus the robust nature of the zero-n gap with respect to input angle persists in the presence of nonlinearity. The Bragg gap soliton disappears for detuning as small as 5 degrees. The reason the Bragg gap soliton exhibits a sensitive dependence on input angle is simply because of its sensitivity to lattice scaling. To achieve a Bragg gap soliton at the new angle one would have to change the incident frequency and increase the input intensity. Beyond a certain input angle, no Bragg gap soliton exists.

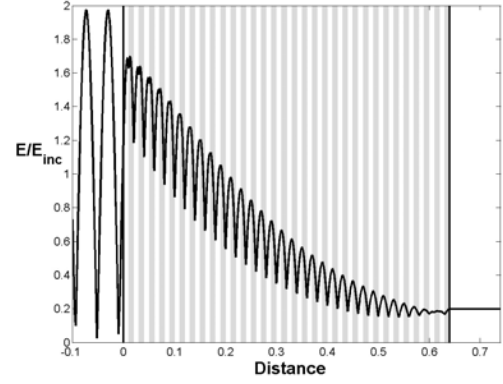

**FIGURE 6.** Spatial distribution of the E field magnitude (normalized w.r.t. $E_{inc}$) with the conditions as in Figure 3, but having a near zero transmittance.

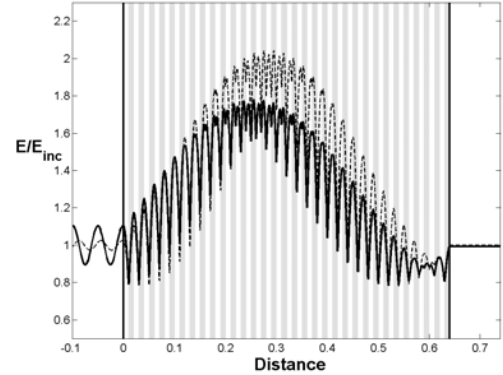

**FIGURE 7**. The Soliton for an incident angle of 30° (solid line) as compared to the one at normal incidence (dotted line). $\gamma I_{inc} = -0.064$ for both cases and other conditions as in Figure 3.

We have also studied the behavior of the zero-n gap soliton with lattice scaling and inclusion of length disorders. In the presence of disorder and scale change, the zero-n gap soliton remains relatively robust in the sense that it can be excited at the same input frequency. However, the intensity needed to achieve exact nonlinear resonance changes somewhat.

In conclusion, we have shown that periodic structures composed of alternating layers of nonlinear positive-index and negative index materials possess a novel gap soliton associated with the zero-average-permittivity band gap. Such zero-n gap solitons retain the exotic features associated with the linear zero-n bandgap.